\documentclass[12pt]{aastex62}
\accepted{ApJ}

\newcommand{\Msun}{$M_\odot$} 
\newcommand{\degs}{\mbox{$^{\circ}$}}

\newcommand{\axaf}{\mbox{\em Chandra\/}}
\newcommand{\einstein}{\mbox{\em Einstein\/}}
\newcommand{\rosat}{\mbox{\em ROSAT\/}}

\usepackage{graphicx}
\usepackage{hyperref}
\usepackage{color}

\usepackage{lineno}
\modulolinenumbers[1]

\shorttitle{X-ray dominated jets}
\shortauthors{Schwartz et al.}

\begin{document}

\title{TWO CANDIDATE HIGH-REDSHIFT X-RAY JETS WITHOUT COINCIDENT RADIO JETS}

\correspondingauthor{Daniel Schwartz}
\email{dschwartz@cfa.harvard.edu}

\author[0000-0001-8252-4753]{D. A. Schwartz}
\affiliation{Smithsonian Astrophysical Observatory,
Cambridge, MA 02138, USA}

\author[0000-0002-0905-7375]{A. Siemiginowska} 
\affiliation{Smithsonian Astrophysical Observatory,
Cambridge, MA 02138, USA} 

\author[0000-0002-4900-928X]{B. Snios} 
\affiliation{Smithsonian Astrophysical Observatory,
Cambridge, MA 02138, USA} 

\author[0000-0002-1516-0336]{D. M. Worrall}
\affiliation{HH Wills Physics Laboratory, University of Bristol,
  Tyndall Avenue, Bristol BS8 1TL, UK} 

\author[0000-0002-1858-277X]{M. Birkinshaw}
\affiliation{HH Wills Physics Laboratory, University of Bristol,
  Tyndall Avenue, Bristol BS8 1TL, UK}

\author[0000-0002-4377-0174]{C. C. Cheung}
\affiliation{Space Science Division, Naval Research Laboratory, Washington, DC
20375-5352, USA}

\author[0000-0002-6492-1293]{H. Marshall}
\affiliation{Kavli Institute for Astrophysics and Space Research,
  Massachusetts Institute of Technology, 77 Massachusetts Avenue, Cambridge, MA 02139, USA}

\author[0000-0003-0216-8053]{G. Migliori}
\affiliation{INAF Istituto di Radioastronomia, via Gobetti 101,
  I-40129 Bologna, Italy}

\author[0000-0002-8960-2942]{J. F. C. Wardle}
\affiliation{Department of Physics, MS 057, Brandeis University,
  Waltham, MA 02454, USA} 
\author{Doug Gobeille}
\affiliation{Department of Physics, University of Rhode Island, 45
  Upper College Rd, Kingston, RI 02881, USA} 


\begin{abstract}
We report the detection of extended X-ray emission from two high-redshift radio quasars. 
  These quasars, J1405+0415 at
  $z$=3.208  and J1610+1811 at $z$=3.118, were observed
  in a \axaf\ 
  snapshot survey selected from a complete sample of the radio-brightest quasars in the
  overlap area of the VLA-FIRST radio survey and the Sloan Digital Sky
  Survey.   The extended X-ray emission is located along the line connecting the core to a radio knot or hotspot, favoring the interpretation of X-ray jets. 
  The inferred rest frame jet X-ray luminosities from 2--30 keV would be of order  10$^{45}$ erg~s$^{-1}$ if emitted isotropically and without relativistic beaming. In the scenario of inverse Compton scattering of the cosmic microwave background (CMB), X-ray jets without a coincident 
  radio counterpart  may be common, and should be readily detectable to redshifts even beyond 3.2  due to the (1+$z$)$^4$ increase of the CMB energy density compensating for the (1+$z$)$^{-4}$
  cosmological diminution of surface brightness. If these can be X-ray confirmed, they would be the second and third examples of quasar X-ray jets without detection of underlying continuous radio jets.

\end{abstract}




\section{Introduction} \label{sec:intro}

Although the first jet from an Active Galactic Nuclei 
(AGN) was discovered as a visible image in a
photograph of M87 \citep{Curtis18}, they have been observed
primarily as radio phenomena
\citep[e.g.,][]{Turland75,Waggett77,Readhead78,Perley79,Bridle84}. Jets
provide a mechanism to explain the morphologies of
extragalactic radio sources and to supply the large energy content
inferred in  lobes of extragalactic radio sources
\citep[e.g.,][]{Rees71,Longair73,Scheuer74,Blandford74}.  Jets transport  energy from the central super-massive black hole to radio lobes, doing work on the external medium, and playing a significant role in the energy budget of  black hole accretion. It is now
recognized that extragalactic jets play an essential role in the
feedback processes that prevent catastrophic cooling-flow collapse 
in clusters of galaxies \citep[e.g.,][]{Fabian00,Fabian12,Hardcastle20}.  \deleted{and that lead to the correlation of the masses of 
super-massive black holes and the central bulge of their galaxies} 

Multi-wavelength data are important for understanding the physics of
these systems.  The \axaf\ X-ray Observatory has enabled X-ray observations
to contribute significantly to the study of the power and morphology
of jets in extragalactic sources \citep{Harris06, Worrall09,
  Schwartz10}.  Using a model-dependent
assumption that the X-rays are generated by inverse Compton (IC)
up-scattering of the CMB radiation \citep{Tavecchio00,Celotti01},  X-rays help estimate the enthalpy flux, often simply
called ``power,'' that does work on the external medium resulting
in feedback. The power carried by kpc-scale jets has generally
been estimated by assessment of the energy deposited into radio lobes
and cocoons \citep{Scheuer74,Rawlings91,Willott99}, or by the energy
required to create cavities observed in the hot intra-cluster or
intra-galactic gas at low redshift \citep{Birzan08}, or by empirical
scaling relations derived from those methods
\citep[e.g.,][]{ODea09,Cavagnolo10,Daly12}. Those are all based on
calorimetry coupled with an estimate of age to give an average power
output.  The IC/CMB interpretation of the X-ray observations offers an alternative method of estimating
power by measurements of the jet itself. 

At high-redshift, several factors favor jets manifesting as X-ray via IC/CMB
rather than as radio emitters.  The radio surface brightness suffers the
cosmological diminution factor $(1+$z$)^{-4}$, while for IC/CMB X-ray
emission this is compensated by the $(1+$z$)^{4}$ increase in the CMB
energy density. Another factor is the $\approx$\,100 times longer
lifetimes of the electrons with energies of order 100 MeV producing X-rays
via IC/CMB, compared to the electrons with order of 10 GeV that give the
GHz synchrotron radiation.  In addition, the observed radio emission is more
diminished by the redshift than  the X-rays since the radio spectrum tends
to steepen at emitted mm-wave frequencies, while the X-rays generally have a flatter
spectrum.  However, searching for new examples of relativistic kpc-scale X-ray jets with \axaf , the only current 
instrument capable of detecting them, is observationally expensive
as only those few percent of radio quasars which are beamed tightly in
our direction are viable candidate systems. The vast majority of 
X-ray jet detections therefore result
from pointed observations of known radio jets.

Serendipitously, \citet{Simionescu16} discovered the first, dramatic example of
an X-ray jet resolved on arcsec-scale and without a corresponding radio jet detection. Here we
present two further candidates, the quasars J1405+0415 at $z=3.208$ \citep{Barthel90}, and J1610+1811 at $z=3.118$ \citep{Osmer94}\footnote{We will refer to these as J1405 and J1610, respectively.}. These systems are distinguished by the absence of a detectable radio jet.  
Preliminary results have been presented in \citet{Schwartz19}.

We adopt ${\rm H}_{0}=67.8\rm\,km\,s^{-1}\,Mpc^{-1}$, $\Omega_{\rm
M}=0.308$ and $\Omega_{\rm \Lambda}=0.692$, \citep{Planck16}, giving scales of 7.8 and 7.7~kpc\,arcsec$^{-1}$
for redshifts of 3.1 and 3.2, respectively.  We use the terminology definitions of \citet{Bridle94} to describe the radio features. Spectral
indices $\alpha$ are defined by flux density S$_{\nu} \propto
\nu^{-\alpha}$. Photon number indices are $\alpha+1$, and corresponding relativistic
electron number spectra are  dN/d$\gamma \propto \gamma^{-(2\alpha+1)}$, where $\gamma$ is the electron Lorentz factor.

\section{The HIGH-REDSHIFT sample}
We carried out an exploratory survey for X-ray jets associated
with high-redshift radio-loud quasars. Our sample was drawn from the complete
survey of \citet{Gobeille11}, \citep[also,][]{,Gobeille14}. 
That survey included the
123 radio-brightest quasars at redshifts greater than 2.5 in the
overlapping region of the VLA-FIRST radio survey \citep{Becker95} and the
Sloan Digital Sky Survey \citep{Abazajian03}. The quasars were 
selected to have a total flux density $>$ 70\,mJy at either 1.4 or 5\,GHz, 
and required to have a spectroscopically-measured redshift. In that sample, 61
systems show resolved radio structure detected with 
1\arcsec\ or finer resolution, and from these we eliminated 30 that were
classified as 
triples since they are not likely to be relativistically beamed in our
direction as is necessary to reveal IC/CMB emission. This left 31 sources with resolved radio structure, for which we
ignored morphological distinctions such as jet, knot, hotspot, or lobe.
We took the 16 with redshifts $z>3$ as most likely to be
detected in short, 10 ks \axaf\ ``snapshot'' observations to look for jets suitable for further followup observations. Of these 16, we did not
re-observe two quasars previously detected, J1430+4204 
\citep{Cheung12} at $z=4.7$, and J1510+5702 \citep{Siemiginowska03,Yuan03} at $z=4.3$. Each quasar has one
well-defined direction given by an  extended radio feature.  \replaced{This paper
presents the two cases of extended X-ray emission without co-spatial radio emission. We will
report separately on the results of our entire survey, including all the quasar core data.}{

This paper
presents two cases of extended X-ray emission without co-spatial radio emission. \emph{A posteriori} the extended emissions are just at the 99.7\% confidence limit, giving  a 0.17\% chance that such a result could arise in 14 trials, and a 4.2\% chance of one such false detection. For comparison, a 99\% confidence result would have allowed a 13\% chance of one such false positive in 14 trials.  In the survey of 14 objects we
   have significant X-ray emission external to the quasar core in
       5 of the sources, including the two reported here, as will be reported by Snios et al.(in preparation), along with the quasar core data. }

The survey observations were done with  ACIS S-3, in the standard 1/4
subarray timed exposure mode, with very faint telemetry format. Pileup was less than 2.5\% for each quasar, and was neglected in the analysis. In the 0.5--7.0 keV X-ray band, background
from the diffuse X-ray sky and from non-X-ray events was very small;
$0.0273\pm0.0027\rm\,counts~arcsec^{-2}$ for J1405, and
$0.0195\pm0.0011\rm\,counts~arcsec^{-2}$ for 
J1610. Therefore  we only used the standard faint mode telemetry processing for reconstructing each individual photon event. All observations had a roll direction
preference such that the extended radio feature would not coincide
with the ACIS readout streak. Each observation was 
approved for 10 ks on target, which would result in a nominal 9.6 ks live time after correction for the 4.88\% dead-time due to the 1/4 subarray readout. \axaf\ observed J1405 (ObsID 20408) for 9.6 ks live time on May 8,
2018, and J1610 (ObsID 20410) for 9.1 ks live time on May 24, 2018.
We used CIAO software version 4.12 \citep{Fruscione2006} and CALDB version 4.9 in data analysis.

To compare with the X-ray images, we obtained new  Karl G. Jansky Very Large Array (VLA) A-array data (program 12B-230). A total of 5.3 minutes of exposure was obtained for each source on 2012 Nov 18 (J1405) and 2012 Dec 08 (J1610). The data were calibrated and imaged with CASA using standard procedures. Each data set used two intermediate-frequency bands (1 GHz bandwidth each) centered at 4.9 and 7.4 GHz, giving an effective center frequency of the resultant images at 6.2 GHz.

\section{Extended X-ray emission}
 The identification as X-ray jets can be justified by the
fact that we have a statistically significant detection of X-ray photons
in an extended linear region defined by the direction of the central source
to radio emission in an external 
knot or hotspot. The region width is defined by the \axaf\ spatial
resolution. The existence of an external radio feature means it must have been, or currently is
being, powered by a jet.  Because of our limited statistics, we cannot claim
we have detected extended X-ray jets according to the formal definition of having a length at least 4
times its width \citep{Bridle84}. 

The dominant background for detecting an arcsec-scale jet is scattered X-rays from the quasar
nucleus. For each object, we fit a power-law spectrum to the quasar.
We use that spectrum in
saotrace-2.0.4\_03\footnote{http://cxc.harvard.edu/cal/Hrma/SAOTrace.html}
to generate rays which are passed to
marx-5.5.0\footnote{https://space.mit.edu/CXC/MARX/} \citep{Davis12} to simulate an
ACIS-S image. We use Marx with the energy dependent sub-pixel event
redistribution (EDSER) algorithm. We run 500 separate simulations with
the actual source flux, observing time, and aspect dithering, in order to accurately simulate the pile-up and
the ACIS readout streak.  The resulting files are merged into a single, simulated 
image of the point-spread function.  The simulated image counts are normalized by the ratio of counts in a 0\farcs95 radius about the quasar, compared to counts in the same region of the simulated image.  The error in this normalization is dominated by the number of counts observed from the quasar, and in turn determines the uncertainty in the  expected number of scattered X-rays in the jet region.
 The 0\farcs95 radius contains a nominal 90\% encircled counts at 1.5 keV, but for the broad quasar spectral distribution our simulations give 83.6\% and 83.7\% respectively for J1405 and J1610. \added{This 0\farcs95 radius is an objective choice of a distance to search for emission external to the quasar core, and is used for J1405. However, since quasar J1610 is 40\% brighter than J1405, using that same distance criterion would mean that our detection sensitivity threshold would be 1.5 times larger. Via simulation we find that for a jet box 1\farcs3 from J1610, the background due to scattered quasar counts is the same for both objects.}

\subsection{J1405+0415}

\begin{figure}
\begin{center}
\includegraphics[width=6.5in]{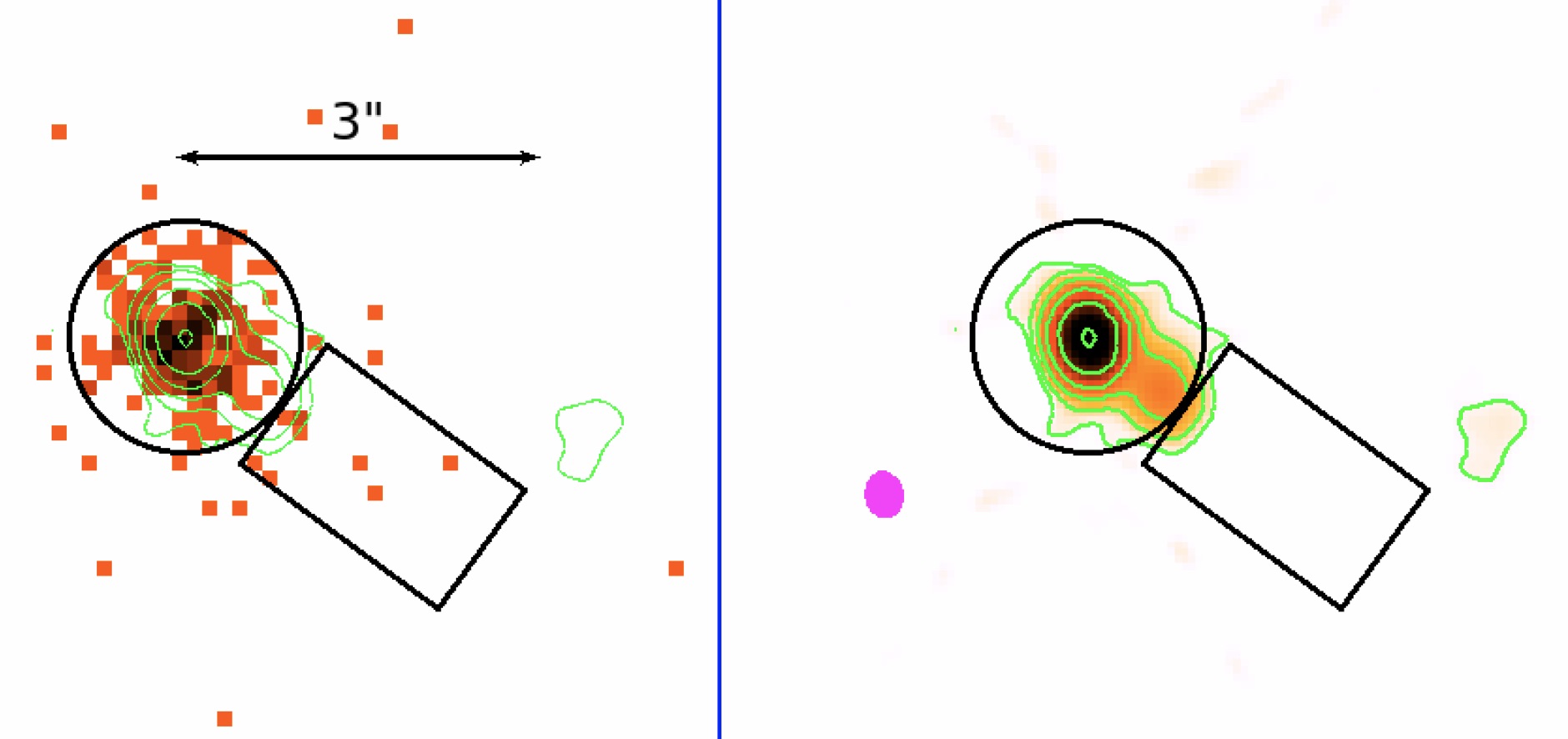}
\caption{Quasar J1405. Left: 0.5--7.0\,keV X-ray data in 0\farcs123  pixels. Maximum counts
  are 13 per pixel. Both panels
  show a 0\farcs95 radius circle about the quasar and the
  2\farcs0$\times$1\farcs5 box used for the extended region we
  designate as the jet.  Right: Our 6.2 GHz VLA  data, showing the
  quasar core, a radio knot at 0\farcs73 from the quasar at position angle
  235\degs, and a faint hotspot about 3\arcsec\ away at position angle
  257\degs. Radio contours are shown in both panels, with the lowest
  level at 0.6 mJy beam$^{-1}$ and  increasing by factors of
  4, with a peak flux density of 694 mJy beam$^{-1}$. Restoring beam shown shaded
  (magenta in on-line version) 
  is
  0\farcs39$\times$0\farcs32 FWHM at 7.5\degs\ position angle. \label{fig:J1405}}
\end{center}
\end{figure}

The flat spectrum radio source PKS 1402+044
\citep{Shimmins75} was identified as a quasi-stellar object by \citet{Condon77}, with a 
redshift $z=3.20$ measured by \citet{Peterson78}. VLBI  observations showed that a
quasar jet was relativistically beamed in our direction on
pc scales with  components at a range of position angles (measured positive east of north)
from $258\degs$ to $318\degs$ at distances 4.4 to 14.5\,milli-arcsec from the
brightest flux density position \citep{Gurvits92}. \citet{Gurvits92} pointed out that these features could be
interpreted as a continuous jet, bending through an apparent angle of
$\approx$90\degs\ and then pointing toward a knot  about 0\farcs8 away at 237\degs\ position angle,  and possibly connecting to faint extended emission
3\farcs3 away at position angle $254\degs$.  They concluded that the
variation of  position angles  was most likely due to a jet closely aligned to our line of sight and  deflected through a relatively small
angle.  Higher resolution VLA, VLBA and VSOP multi-wavelength observations in March 1998, and January and October 2004 
by \citet{Yang08} confirmed the morphological structure and 
resolved further pc-scale components with a range of position angles $232.2\degs$ to
$334.4\degs$. They determined that several of the core components had
brightness temperatures near and above $10^{12}$\,K, indicating
relativistic motion. For the innermost components they deduced a
Doppler factor $\delta > 23$, and an upper limit to the angle to our
line of sight of 1\degs . They fit the GHz spectra of the VLA-observed arcsecond-scale features to power-laws. Renormalizing to our central frequency, they report  670($\nu$/$6.2 _{\rm GHz}$)$^{-0.09}$ mJy for the
core, 34($\nu$/$6.2 _{\rm GHz}$)$^{-0.91}$ mJy for the knot, and
2.9($\nu$/$6.2 _{\rm GHz}$)$^{-1.66}$ mJy for the lobe in the 1.4 to 15.9 GHz range, all in agreement with our measurements as reported below. 


Using  the
\emph{Einstein} Observatory, \citet{Zamorani81} first detected X-rays from the quasar, reporting a flux of
$(0.8\pm0.4)\times10^{-13}\rm\,erg\,cm^{-2}\,s^{-1}$ in the 0.5--4.5 keV
band. \citet{Brinkmann97} reported a flux of $(2\pm1.2)\times10^{-13}\rm\,erg\,cm^{-2}\,s^{-1}$ in the 0.1--2.4\,keV band,
based on pointed \emph{ROSAT} observations. Considering  the uncertainties and the bandwidth differences 
the X-ray flux may have been constant. Both those telescopes had
$\approx$\,5\arcsec\ resolution and therefore could not have resolved 
any small-scale extent.

 Figure~\ref{fig:J1405} shows our 0.5--7.0 keV X-ray and
6.2 GHz radio data for J1405. 
 We shifted the \axaf\ image by 0\farcs 57 so that the quasar centroid
coincided with the radio position at 14$^h$05$^m$01.12$^s$ +4\degs\,15\arcmin\,35\farcs8.
We take the quasar region counts inside a 0\farcs95
radius to determine the X-ray spectrum.  The 264 quasar counts are 
fit to a power-law with fixed n$_H$=2.19$\times$10$^{20}$ H-atoms
cm$^{-2}$, using the Cash statistic in CIAO 4.12 Sherpa version 1 \citep{Freeman2001}. The best fit gives a 
spectral index $\alpha=0.38\pm 0.12$, for which the incident
0.5--7.0 keV flux corrected for Galactic absorption 
is  $(4.0\pm0.4)\times10^{-13}\rm\,erg\,cm^{-2}\,s^{-1}$, 
corresponding to a rest-frame luminosity of
$(3.8\pm0.4)\times10^{46}\rm\,erg\,s^{-1}$ in the 2.1--29.4\,keV band. The flux is consistent with the \rosat\  flux, but a factor of two higher than that measured with \einstein , all extrapolated to the same energy range.

The rectangle in Figure~\ref{fig:J1405} shows the region we take 
for the X-ray jet. It is extended
2\arcsec\ beyond the quasar region,  parallel to and straddling the line
from the radio core through the knot at 0\farcs7, as shown in both panels of
Figure~\ref{fig:J1405}. Figure 3a of \citet{Yang08} indicates that the radio jet changes direction about 2\farcs6 from the quasar, bending  toward the lobe and hotspot at position angle 254\degs , but we do not have sufficient signal to investigate this in X-rays. \added{We therefore ended the jet box at this distance.} We take a width of 1\farcs5, which is the fit of the FWHM to the 6900 counts in the readout streak at position angles 147\degs\ and 327\degs\  of the simulated image. The X-ray image contains 9 counts in this region, while the simulated X-ray
image gives 
1100 counts in this box. We scale the latter number by the ratio of 264
counts in the quasar circle to 118472 counts in the same circle of the
simulated image, predicting that $2.45\pm0.17$ counts from the quasar will
scatter into the region taken for the jet. Taking 2$\sigma$ above the predicted counts, and  an additional 0.082
counts from the background, Poisson statistics gives a probability of 0.29\% for the
null hypothesis of zero extended X-ray emission. Assuming the same
spectrum as the quasar, the net jet flux would be $1.\times10^{-14}\rm\,erg\,cm^{-2}\,s^{-1}$ with an uncertainty of a factor of 2. The rest-frame 2.1--29.4 keV
luminosity, if the radiation were unbeamed and isotropic, would be
$9\times10^{44}\rm\,erg\,s^{-1}$. \added{(If we had started the jet box 1\farcs1 from the quasar, we would have had only 7 photons, but a lower background of 2.22 counts, and the null hypothesis probability would increase to 0.8\%.)}

\begin{figure}
\includegraphics[width=6.in]{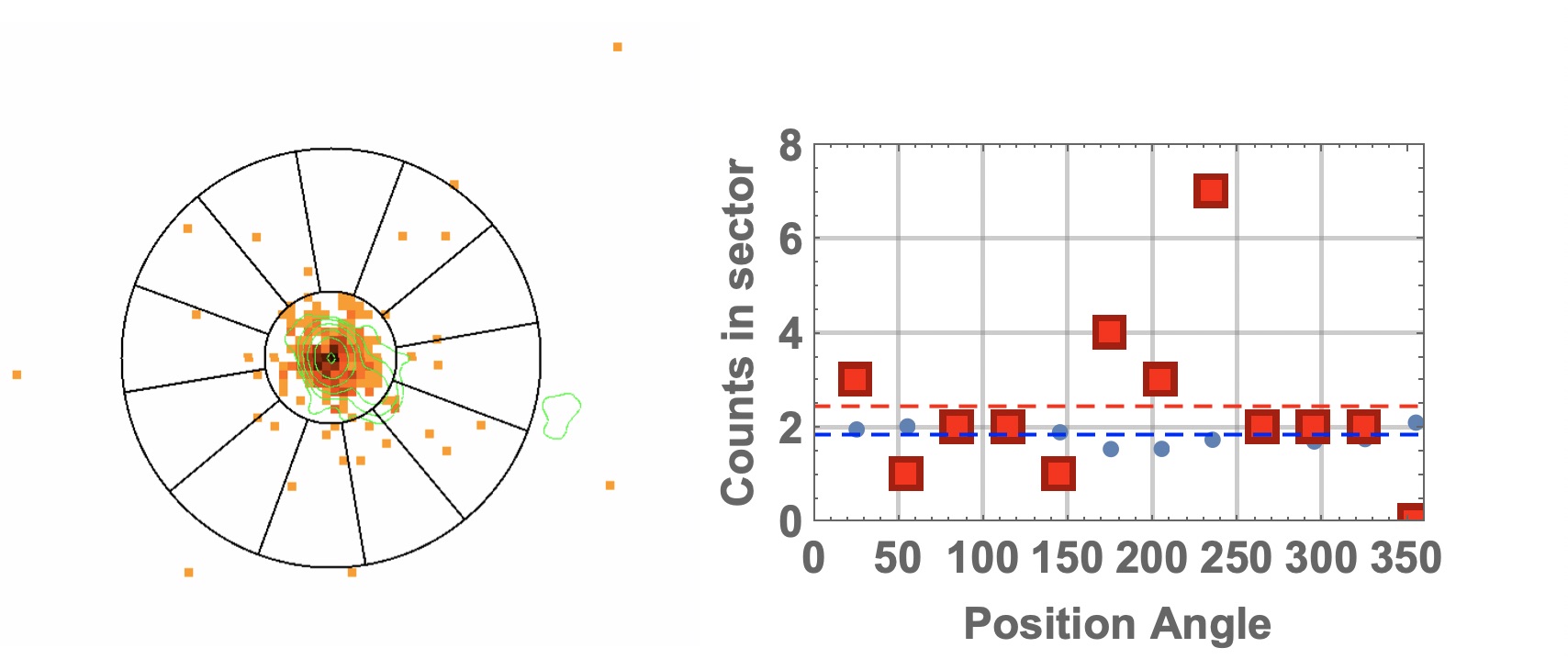}
\caption{Left: The J1405 X-ray data showing twelve 30\degs\ sectors emphasizing  the
  excess X-ray counts from Fig.~\ref{fig:J1405} in the direction of
  the radio 
  extension. The annulus is 0\farcs95 to 3\arcsec\ from the
  X-ray core, \added{i.e., identical to the length of the jet box in Fig 1}. Right: Counts per 30\degs\ sector in the annulus as a function of position angle measured counter-clockwise from North. Dots (blue in the on-line version) are from
the SAOTrace/Marx simulation scaled to
the  number of counts from the quasar.
Squares (red in the on-line version) are the data.
Counts in the sector at 235\degs\ are significantly 
above the average (dashed lines). The readout
streak would occur at position angles 147\degs and 327\degs\ and is undetectable due to the  low count rate of the quasar and the short exposure time. \label{fig:1405panda}}
\end{figure}

The significance of the extended X-rays can also be assessed by the azimuthal distribution of counts. Figure~\ref{fig:1405panda} divides the
X-ray image into an annulus from 0\farcs95 to 3\arcsec\ and into
30\degs\ sectors, centering one sector on the position angle 235\degs defined by the direction to the knot at 0\farcs7. The right hand panel of Figure~\ref{fig:1405panda} plots the counts in
each sector of the annulus. The average of the data in the 12 sectors
is 2.41 counts per sector, and the probability of getting 7 counts in the sector with
the radio knot \added{by chance is only} 1.2\%. This calculation gives a higher chance
probability of a spurious result because the sector shape is narrower
than the \axaf\ telescope resolution and does not capture all the jet
counts. Based on the simulated data, the average counts per
sector is only 1.85, and the probability of observing 7 counts in the sector with
the radio knot would be  0.30\%.

The rectangular region used to assess the extended X-ray emission contains a 6.2 GHz flux density of 2.4 mJy from the knot at 0\farcs7 and its extension. We use that flux density in computing the lower limit X-ray to radio emission ratio. Beyond those radio contours and within the remainder of the X-ray jet region in Figure~\ref{fig:J1405} we derive an upper limit by computing the rms Jy beam$^{-1}$ for the 0\farcs06$\times$0\farcs06 pixels. This gives 0.134 mJy beam$^{-1}$. Multiplying by  3 times the square root of the number of beam areas in the rectangle gives an upper  limit of 1.3 mJy for the 6.2 
GHz radio emission in the region beyond the radio contours. Our measured flux densities for the core, knot at 0\farcs7 distance, and lobe at 3\farcs3 distance are 650 mJy, 25.5 mJy, and 1.7 mJy, respectively. These fall within the error bars of the measurements shown in Figure 5 of \citet{Yang08}.

\begin{figure}[t]
\centering
\includegraphics[angle=-90.,width=4.5in]{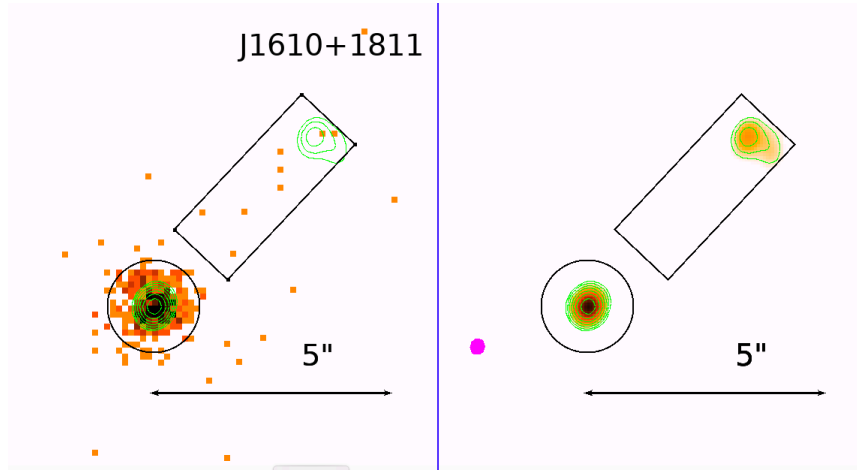}
\caption{Quasar J1610. Left panel: Our 0.5 to 7 keV \axaf\ X-ray data binned in 0\farcs123
  pixels. Maximum counts
  are 13 per pixel. The 0\farcs95 radius circle centered on the quasar contains
  370 photons. The 3\farcs8$\times$1\farcs5 rectangular region taken
  for the jet parallels the line from the quasar to the hotspot in the
  radio image. Right panel: Our 6.2 GHz VLA image. Radio contours are
  shown in both panels, and increase from  0.5 mJy beam$^{-1}$ by factors of 2, with a
  peak flux density of 66.3 mJy beam$^{-1}$ at the quasar. Restoring
  beam of 0\farcs17$\times$0\farcs15 at position angle 343.45\degs\
  is shown shaded (magenta in on-line version). \label{fig:1610}}
\end{figure}

\subsection{J1610+1811}

TXS 1607+183  was discovered with a 365\,MHz flux
density of 415\,mJy \citep{Douglas80, Douglas96} in the Texas Survey,
and in the MIT Green Bank survey with a flux density of 165\,mJy at
4.8\,GHz \citep{Bennett86}. The spectral index $\alpha=0.37$ between
those two frequencies classified the object as a flat spectrum radio
quasar. \citet{Osmer94} reported a redshift $z$=3.118. The
\emph{ROSAT} all sky survey faint source catalog \citep{Voges00} lists
this quasar with $0.0162\pm0.0072\rm\,counts\,s^{-1}$ from 0.1--2.4\,keV.  That rate would correspond to a flux of (4.4$\pm$2)$\times$10$^{-13}$\,erg cm$^{-2}$ s$^{-1}$ from 0.5 to 7 keV according to the WebPIMMS\footnote{https://heasarc.gsfc.nasa.gov/cgi-bin/Tools/w3pimms/w3pimms.pl} \citep{Mukai93} count rate converter.

 We shifted the \axaf\ image by 0\farcs28 so that the
quasar centroid coincided with the radio position at 16$^h$10$^m$05.29$^s$
+18\degs\,11\arcmin\,43\farcs4.  Figure~\ref{fig:1610} shows our 0.5--7.0\,keV and
6.2 GHz images of J1610. In the radio image, a 4\farcs76 long
line from the quasar position to the center of a radio hotspot
defines the direction of an expected jet at $317\degs$ position
angle.  \citet{Bourda11} find a 2 milli-arcsec long VLBI jet at
essentially the same position angle, $316\degs$. Excluding the core and the lobe, we measure 
a 3$\sigma$ limit of about 0.34 mJy
to the total 6.2 GHz flux density in the region between. We measure S$_{6.2}$=67 mJy for the quasar core, and S$_{6.2}$=7.6 mJy from the NW lobe.

We fit the 370 X-ray counts inside a 0\farcs95 circle about the quasar to a
power-law with fixed Galactic column density of 3.73$\times$10$^{20}$ H-atoms
cm$^{-2}$. The maximum likelihood Cash-statistic gives a best fit spectral index $\alpha=0.61\pm0.10$. The measured 0.5--7.0 keV flux of
$(5.8\pm0.5)\times10^{-13}\rm\,erg\,cm^{-2}\,s^{-1}$ corrected for Galactic absorption implies a rest-frame
luminosity of $(5.5\pm0.5)\times10^{46}\rm\,erg\,s^{-1}$ in the 2.1--28.8\,keV band. 

\begin{figure}
\includegraphics[width=6.in]{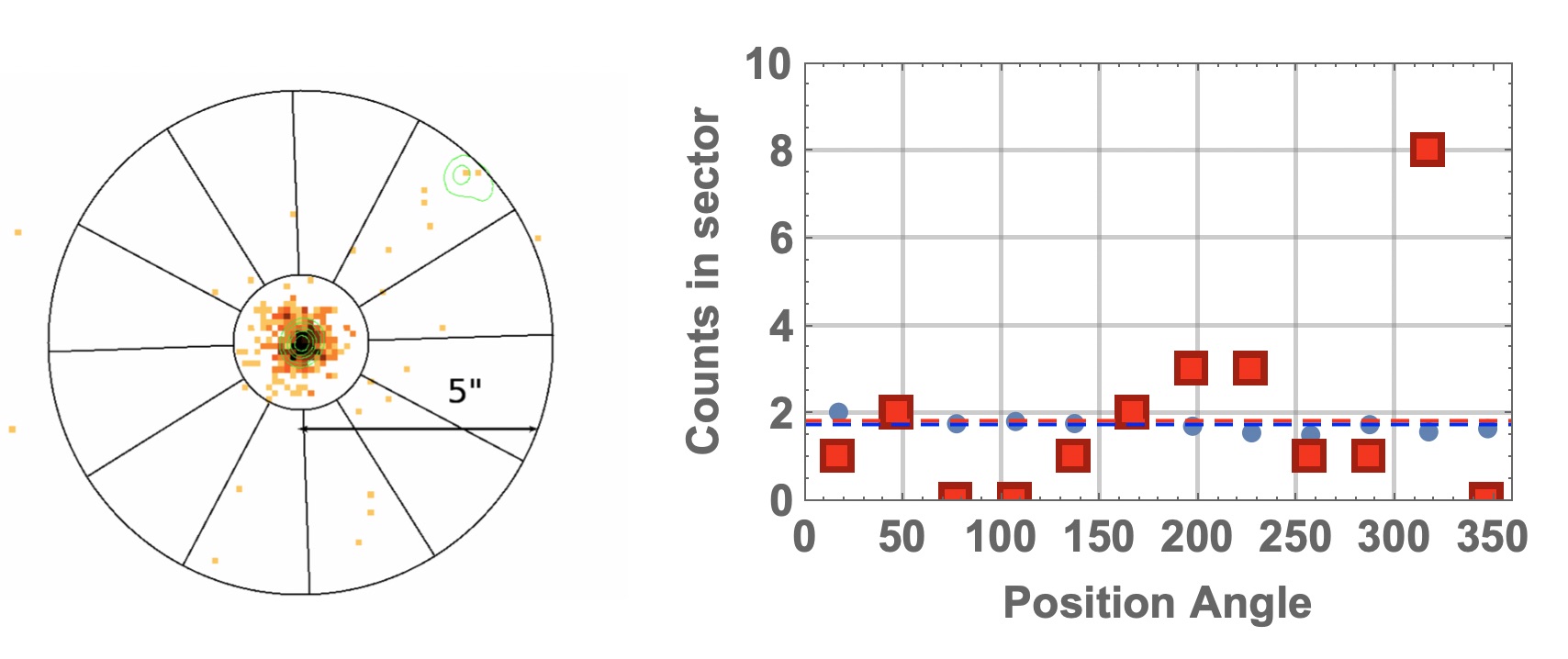}
\caption{Left: The contours present our 6.2 GHz
  measurements of J1610.  The sectors  emphasize the
  excess X-ray counts from Fig.~\ref{fig:1610} in the direction of
  the radio 
  extension. The annulus  is 1\farcs4 to 5\farcs2 from the
  X-ray core, \added{i.e., identical to the length of the jet box in Fig 3}. Right: Counts between 1\farcs4 and 5\farcs2 from the quasar, per 30\degs\ sector vs. position angle measured counter-clockwise from North. Dots (blue in the on-line version) are from
the SAOTrace/Marx simulation scaled to the number of counts from the quasar.
Squares, (red in the on-line version) are the data.
The 8 counts in the sector at 317\degs\  are significantly 
above the average (dashed lines) of 1.83 counts per sector. The readout
streak would occur at position angles 0.2\degs and 180.2\degs\ and is undetectable due to the relatively low count rate of the quasar and the short observation time. \label{fig:1610panda}}
\end{figure}

The quasar X-ray core of J1610 is 40\% brighter than that of J1405. This prevents testing  for a jet at
the same 0\farcs95 distance from the quasar because of the higher density of scattered X-rays. Figure~\ref{fig:1610} (left
panel)
shows  a region 3\farcs8 long and 
1\farcs5 wide along the position angle to the radio hotspot and displaced 1\farcs3 
from the quasar.  There are
8 X-ray photons in this region. The simulation of the quasar  predicts
that 1.88$\pm$0.12  counts are expected to be scattered X-rays from the
quasar core. Taking 2$\sigma$ above the expected scattered counts and adding the detector background of 0.11 counts, the chance
of observing 8 under the null hypothesis of no extended X-ray emission
is 0.22\%. If we shrink the X-ray jet region to exclude the radio lobe, there are 6 X-ray photons, and a 1.6\% probability that there is no extended emission. Sector analyses similar to that done for J1405 were carried out. The result using the full X-ray region indicated in Figure~\ref{fig:1610} is shown in Figure~\ref{fig:1610panda}, and gives a 0.06\% probability that there is no X-ray extension. A sector analysis extending only to 4\farcs5 from the quasar to exclude the radio lobe, gives a 0.6\% probability for observing the 6 counts when expecting 1.53 counts per sector, the average of all twelve sectors. Assuming the same spectrum as the quasar, the 6 net counts
in the jet convert to  a flux of $0.9\times10^{-14}\rm\,erg\,cm^{-2}\,s^{-1}$
and a rest frame 2.1--28.8\,keV luminosity of $9\times10^{44}\rm\,erg\,s^{-1}$,
if isotropic and without relativistic beaming.  The uncertainties are a factor of 2. 

\section{The  X-ray dominated jets}

The IC/CMB mechanism offers the simplest model of the jet if
the broad-band emission is attributed to a single spectrum of
relativistic electrons
\citep[e.g.,][]{Tavecchio00,Celotti01,Siemiginowska02,Sambruna02,Sambruna04,Sambruna06,Marshall05,Schwartz05,Marshall11,Marshall18,Schwartz06a,Schwartz06b,Worrall09,Perlman11,Massaro11}.
For the low redshift, $z\le2.1$, jets modeled in those references it
is also required that there is bulk relativistic motion with Lorentz 
factor $\Gamma \ge 3$ with respect to
the co-moving frame of the parent quasar. Bulk relativistic motion is
generally inferred for powerful, one-sided quasar jets. 

However, there is clear evidence for multiple non-thermal electron
populations in jets with complex X-ray morphology, e.g. in 3C 273 \citep{Jester06} at $z$=0.158, PKS\,1127-145 at $z$=1.1 \citep{Siemiginowska07}, and PKS 0637-752
\citep{Meyer15} at $z$=0.653. Upper limits on the 0.1--100 GeV $\gamma$-rays from
\emph{Fermi} observations have been used to argue that IC/CMB cannot
give rise to X-ray emission in the jet of PKS 0637-752 \citep{Meyer17}
and four other jets \citep{Breiding17} at redshifts 0.48--1.045, including one that had been
modeled as IC/CMB. Nonetheless, at redshifts $>$3, IC/CMB will be the
predominant loss mechanism whenever the magnetic field strength is
less than 52 $\Gamma\, \mu$Gauss, where $\Gamma$ is the bulk Lorentz factor of the
jet, since the energy density of the CMB at redshift $z$ is equivalent
to a magnetic field of $3.24(1+z)^2\,\mu$Gauss.  \cite{McKeough2016} noticed a possible increase in the X-ray to radio energy flux ratio at $z>3$ in a very small sample of X-ray jets. The increasing
dominance of IC/CMB at large redshifts has been noted for the limits
it may impose on the sizes and ages of FR II radio sources
\citep{Scheuer77,Blundell99}. Those limits will affect survey completeness as well as
bias  correlations of radio source properties
\citep{Blundell99} and  the evolution of black hole activity with
redshift \citep{Simionescu16}. X-ray searches for bright jets which
are faint in radio are important to assess the extent of such biases.

There are some interesting similarities and differences among
J0730+4049, the first X-ray jet discovered without a radio jet at 1.4 GHz \citep{Simionescu16}, the two quasars presented here, and the two other, previously observed, quasars in our defined sample. Observed properties are
summarized in Table~\ref{tab:comparison}. Data for J0730 are from \citet{Simionescu16}, for J1430 from \citet{Cheung12} and for J1510 from \citet{Siemiginowska03}, \citet{Cheung04} and \citet{Cheung05}. Data for J1405 and J1610 are from this paper.

Surface brightness is calculated arbitrarily assuming a 0\farcs5 width,
approximately 4\,kpc, for each jet. The 0\farcs5 is the nominal FWHM of the \axaf\ telescope resolution and comparable to numbers found in the jet of PKS J1421-0643 at $z=3.69$ \citep{Worrall20}.
The roughly similar surface
brightness of these objects is consistent with the expectation from the IC/CMB scenario
\citep{Schwartz02}. While the J0730 jet is similar in surface brightness to the others, the J0730 quasar is an order of magnitude less luminous than the others in this table, and also less than those in Table 6 of \citet{Worrall20}. J1405 and J1610, as well as  J1430+4204 and
J1510+5702, have nearly the
median 2\% jet/core ratio in X-rays found by \citet{Marshall18}. The latter
comparison is surprising. A higher ratio would be expected  if all low
redshift jets were dominated by IC/CMB, but the present result is consistent with the
conclusion of \citet{Marshall11} that the X-ray to radio flux density
ratio does not follow the expected $(1+z)^4$ dependence.  However,
with the expectation that X-ray jets at $z>3$ are due to IC/CMB, the
similarity of the jet/core X-ray ratios to the objects in the \citet{Marshall18} survey  is unexplained. 
It could be that the quasar core X-ray emission is also dominated  by a beamed component as suggested by \citet{Worrall87}. Note that if these jets are beamed in our direction at less than a 10\degs\ angle, then at least the innermost 42 kpc of the jet will appear as part of the X-ray core. If the quasar core X-rays are isotropic, a lower jet to core ratio may be due to larger angles to our line of sight. 
Deeper radio observations and longer \axaf\ observations are necessary to reveal
the spectrum and spatial structure of the extended X-ray and radio emission 
of these systems as well as to
understand the emission mechanism and relation to the super-massive
black hole powering the quasar.

\begin{deluxetable}{cccccccccc}
\tabletypesize{\scriptsize}
\tablecaption{Comparison of high-redshift X-ray jets \label{tab:comparison}}
\tablewidth{0pt}
\tablehead{
 &  & & & &\colhead{X-ray Jet}&\colhead{X-ray Jet} &\colhead{X-ray Jet}&\colhead{}& \colhead{X-ray}\\ 
 &  & \colhead{BH mass\tablenotemark{a}} & \colhead{Live time}&
\colhead{Net Jet counts}& \colhead{Flux\tablenotemark{b} } &
\colhead{Length} & 
  \colhead{ Surface }&\colhead{X-ray/Radio}& \colhead{jet/quasar}\\
\colhead{Name} &\colhead{Redshift}& \colhead{10$^9$\Msun} &
\colhead{ks}&\colhead{0.5--7.0 keV}&
&\colhead{(arcsec)}&\colhead{Brightness\tablenotemark{b}}&\colhead{ratio\tablenotemark{c}}
& \colhead{ratio\tablenotemark{d}}}
\startdata
J0730+4049	&	2.50&0.23&19.&38	&	2.7	&12	&
0.45& $>$73 &0.18	\\
J1405+0415	&3.208&0.87&9.6&6.5	&	1.	&2	&
1.&$>$12&0.025\\
J1430+4204	&4.72&1\tablenotemark{e}&10.6 &20.3	&	1.3	&	3&	0.9&205&0.009 \\
J1510+5702	&4.30& 0.32 &89 &123.5	&	0.76	&	3&	0.5&285& 0.03 \\
J1610+1811	&3.118&10&9.1&6	&	0.9	&	3.8	&
 0.5 &$>4$&0.016\\
\enddata
\tablenotetext{a}{\citet{Shen11}}
\tablenotetext{b}{Flux (10$^{-14}$ erg cm$^{-2}$
s,$^{-1}$) and Surface Brightness (10$^{-14}$ erg cm$^{-2}$
s$^{-1}$ arcsec$^{-2}$) are from 0.5 to 7 keV .}
\tablenotetext{c}{$\nu$F$_{\nu}$ at 1 keV X-ray divided by $\nu$F$_{\nu}$ at 6.2 GHz}
\tablenotetext{d}{Ratio of X-ray counts}
\tablenotetext{e}{\cite{Fabian99}}
\end{deluxetable}




\vspace{5mm}

D.A.S., A.S., and B.S. acknowledge support of NASA contract NAS8-03060
to SAO, and grant GO8-19077X from the CXC. Work by C.C.C. at the Naval Research Laboratory is supported by NASA DPR
S-15633-Y. This research made use of
the NASA Astrophysics Data System and the NASA/IPAC Extragalactic
Database (NED), which is operated by the Jet Propulsion Laboratory,
California Institute of Technology, under contract with the National
Aeronautics and Space Administration. The National Radio Astronomy Observatory is a facility of the National Science Foundation operated under cooperative agreement by Associated Universities, Inc. DAS thanks Daniel Reese for a script used to generate the quasar simulations.

\vspace{5mm}
\facilities{Chandra(ACIS), VLA}

\software{ciao-4.12, SAOTrace-2.0.4\_03, Marx-5.5.0, PIMMS v4.11, CASA 4.7.1-REL
  (r39339)}



\end{document}